# Automatic Optimization of Hardware Accelerators for Image Processing

Oliver Reiche, Konrad Häublein, Marc Reichenbach, Frank Hannig, Jürgen Teich, and Dietmar Fey
Department of Computer Science
Friedrich-Alexander University Erlangen-Nürnberg (FAU), Germany
Email: {oliver.reiche, konrad.haeublein, marc.reichenbach, hannig, teich, dietmar.fey}@cs.fau.de

*Abstract*—In the domain of image processing, often real-time constraints are required. In particular, in safety-critical applications, such as X-ray computed tomography in medical imaging or advanced driver assistance systems in the automotive domain, timing is of utmost importance. A common approach to maintain real-time capabilities of compute-intensive applications is to offload those computations to dedicated accelerator hardware, such as Field Programmable Gate Arrays (FPGAs). Programming such architectures is a challenging task, with respect to the typical FPGA-specific design criteria: Achievable overall algorithm latency and resource usage of FPGA primitives (BRAM, FF, LUT, and DSP). High-Level Synthesis (HLS) dramatically simplifies this task by enabling the description of algorithms in well-known higher languages (C/C++) and its automatic synthesis that can be accomplished by HLS tools. However, algorithm developers still need expert knowledge about the target architecture, in order to achieve satisfying results. Therefore, in previous work, we have shown that elevating the description of image algorithms to an even higher abstraction level, by using a Domain-Specific Language (DSL), can significantly cut down the complexity for designing such algorithms for FPGAs. To give the developer even more control over the common trade-off, latency vs. resource usage, we will present an automatic optimization process where these criteria are analyzed and fed back to the DSL compiler, in order to generate code that is closer to the desired design specifications. Finally, we generate code for stereo block matching algorithms and compare it with handwritten implementations to quantify the quality of our results.

## I. Introduction and Related Work

Real-time image processing is an important task in many application domains. For example autonomous driving or process control need embedded devices for their calculation devices to meet area and energy constraints. Therefore, the traditional way, that an image sensor just captures image data and transfers it to a processing system is not feasible. Rather, the data has to be processed where the information is acquired, which means in or near the image sensor. This leads to a new class of devices, called *smart cameras*. IEEE describes such smart sensor as follows "A transducer that provides functions beyond those necessary for generating a correct representation of a sensed or controlled quantity. This functionality typically implies the integration of the transducer into applications in a networked environment." [1]

One of the first smart cameras was developed by the group of Wolf [2]. They used a *Trimedia CPU* for image preprocessing tasks. To achieve higher frame rates, they proposed to heavily use SIMD[1] instructions. Other approaches, described in [3], use Digital Signal Processors (DSPs) to achieve a very high computing power. To further increase performance, they build a scalable system that consists of up to 10 DSPs for parallel processing.

[1] SIMD: *Single Instruction, Multiple Data*, according to M. Flynn's taxonomy

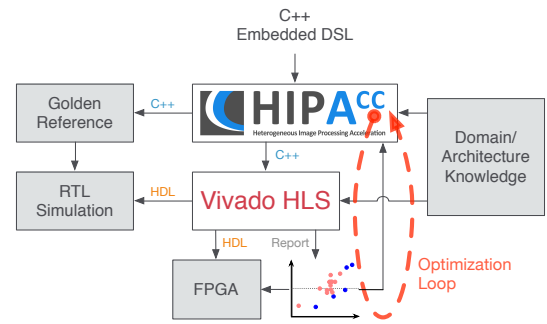

Figure 1. HIPA[cc] design flow with Vivado HLS.

Even more customized architectures have been developed. For example, in [4], a dedicated integrated circuit was developed to speed up image processing within smart cameras. A good survey of smart camera approaches is provided in [5].

With the emerging technology of FPGAs, these devices have been quickly used for the design of smart camera systems. One big advantage is the number of parallel processing units, which can be instantiated in FPGAs as 1D or 2D arrays, since image processing algorithms are in general well parallelizeable [6]. Therefore many new architectures were created on the basis of FPGAs in the past years. While the individual components (e.g., DSP, CPU, FPGA) are well known, a complete design flow how to use this architectures, especially in the domain of image processing is still an open question. Also the combination of such devices to utilize the architectural peculiarities, as described in [7], known as heterogeneous systems, is not completely solved now.

It is well known that application-specific hard- and software will give the highest performance and/or lowest resource utilization. On the other hand, application-specific development is a time consuming and error prone task. Therefore, other approaches were created to describe image processing algorithms in a more abstract way and to perform an automatic derivation.

Schmid et al. proposed in [8] a pipeline design for range image preprocessing on FPGAs. Here, several filters for compensating sensor deficiencies (e.g., noise and pixel defects) were designed by using the HLS framework PARO [9] and evaluated in an experimental setup, consisting of a Microsoft Kinect and Xilinx Virtex-6 LX240T FPGA. Whereas we consider stereo cameras, the authors in [8] mainly focus on different sensor technologies, such as *structured light* and *Time-of-Flight (ToF)*.

Another approach is taken by the HIPA[cc] framework [10] to generate code for FPGA HLS. HIPA[cc] is a publicly available framework[2] for the automatic code generation of image processing algorithms for Graphics Processing Unit (GPU) accelerators. Starting from a C++ embedded DSL, HIPA[cc] delivers tailored code variants for different target architectures, significantly improving the programmer's productivity [11]. Recently, HIPA[cc] was extended to also be able to generate C++ code for the C-based HLS tool Vivado HLS [12], even capable of handling complex multiresolution applications [13]. The design flow of the approach is depicted in Figure 1.

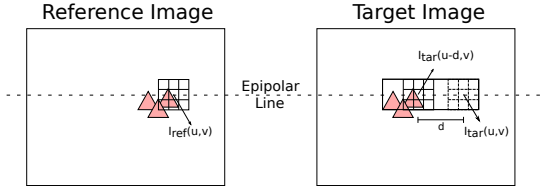

Figure 2. Block matching principle. A window in the reference images, with the center point at $I_{ref}(u,v)$, gets compared to several sub windows in the target image. The distance $d$ defines how far the candidate center point $I_{tar}(u-d,v)$ is shifted in relation to $I_{ref}(u,v)$ on the epipolar line. $d$ correlates with the object distance.

In this paper we present a new extension for automatic optimization, considering given FPGA-specific constraints, to an existing image processing framework. Furthermore, a comparison between a handwritten application-specific architecture development and the utilization of an image processing framework for FPGA targets for smart cameras is made. To make a fair comparison, we are choosing block matching algorithms for the calculation of 3D images from stereo camera systems. Those algorithms are discussed in Section II. In Section III we present the framework and it's new extension for automatic optimization. Finally, we evaluate the results of the optimization process and compare our generated HLS code with highly efficient handwritten implementations in Section IV.

## II. BLOCK MATCHING FOR STEREO CAMERAS

One of the biggest challenges in stereo vision is finding correspondences in pairs of stereo images. This way, the distances of objects in a captured scene can be calculated and saved in a depth or disparity map. Along many techniques solving this issue, stereo block matching is widely used, due to its straight forward procedure. In stereo block matching one image must be defined as reference image, while the other gets determined as target image. It is assumed that each object within a local region of the reference image can be found along the common epipolar line in the target as illustrated in Figure 2.

A local region is defined as a squared block or window with a static pixel range (e.g. $3 \times 3$). The search for correspondences gets further limited by setting of the maximum disparity, illustrated by rectangular block in the target image of Figure 2. Evaluating how similar the reference block is to a sub window block of target image is done by a cost function, which ranks each compared sub window. Common cost functions are Sum of Absolute Differences (SAD) and the Census difference, which are explained in Figure 3.

Lower ranked sub window blocks indicate a closer match to the reference block. Therefore, after each sub window of the target window was compared, the lowest cost function value must be found. For the closest match the found distance $d$ correlates with the distance of the viewed object. High values of $d$ indicate low distance from the image view to the object. This entire process needs to be repeated for every pixel of the reference image.

[2]http://hipacc-lang.org

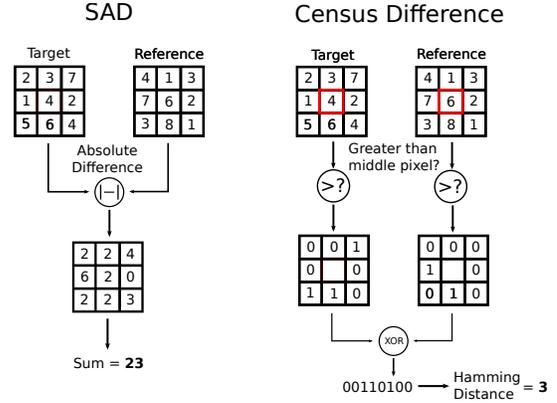

Figure 3. Procedure of cost functions. Left: SAD - Sum of Absolute Difference between values of the masks pixel position is computed. In the final step all window values are summed up. Right: Census Difference - Each pixel is set into relation to the middle pixel of the reference and the target mask (red square). A binary vector can be formed from the windows. After applying a XOR operation, the Hamming distance can be determined.

## III. CODE GENERATION FOR FPGAS

### A. Heterogeneous Image Processing Acceleration Framework

The HIPA[cc] framework consists of a DSL for image processing that is embedded into C++ and a source-to-source compiler. Exploiting the compiler, image filter descriptions written in DSL code can be translated into multiple target languages such as Compute Unified Device Architecture (CUDA), Open Computing Language (OpenCL), Renderscript as used on Android, and C++ code that can be further processed by Vivado HLS [12]. In the following, we will use the Gaussian blur filter as an example to briefly describe properties of the DSL and show how code generation is accomplished.

*1) Domain-Specific Language:* Embedded DSL code is written by using C++ template classes provided by the HIPA[cc] framework. The most essential C++ template classes for writing 2D image processing DSL codes are: (a) an `Image`, which represents the data storage for pixel values; (b) an `IterationSpace` defining the Region of Interest (ROI) for operating on the output image; (c) an `Accessor` defining the ROI of the input image and enabling filtering modes (e.g., nearest neighbor, bilinear interpolation, etc.) on mismatch of input and output region sizes; (d) a `Kernel` specifying the compute function executed by multiple threads, each spawned for a single iteration space point; (e) a `Domain`, which defines the iteration space of a sliding window within each kernel; and (f) a `Mask`, which is a more specific version of the `Domain`, additionally providing filter coefficients for that window. Image accesses within the kernel description are accomplished by providing relative coordinates. To avoid out-of-bound accesses, kernels can further be instructed to implement a certain boundary handling (e.g., clamp, mirror, repeat) by specifying an instance of the class `BoundaryCondition`.

To describe the execution of a Gaussian blur filter, we need to define a `Mask` and load the Gaussian coefficients, defined as

```
1 // input image
2 const int width = 512, height = 512;
3 uchar *image = (uchar*)read_image(width, height, "input.pgm");
4
5 // Gaussian coefficients
6 const float coef[3][3] = { { 0.0625f, 0.1250f, 0.0625f },
7                            { 0.1250f, 0.2500f, 0.1250f },
8                            { 0.0625f, 0.1250f, 0.0625f } };
9
10 Mask<float> mask(coef);
11 Image<uchar> in(width, height);
12 Image<uchar> out(width, height);
13
14 // load image data
15 in = image;
16
17 // reading from in with clamping as boundary condition
18 BoundaryCondition<uchar> bound(in, mask, BOUNDARY_CLAMP);
19 Accessor<uchar> acc(bound);
20
21 // output image
22 IterationSpace<uchar> iter(out);
23
24 // define kernel
25 Gaussian filter(iter, acc, mask);
26
27 // execute kernel
28 filter.execute();
```

Listing 1. Example code for the Gaussian blur filer with kernel size $3 \times 3$.

```
1 class Gaussian : public Kernel<uchar> {
2     // ...
3     void kernel() {
4         float sum = convolve(mask, HipaccSUM, [&] () -> float {
5             return mask() * input(mask);
6         });
7         output() = (uchar)(sum + 0.5f);
8     }
9 };
```

Listing 2. Kernel for the Gaussian blur filter.

constants, see Listing 1 (lines 6–10). It is further necessary to create an input and an output image for storing pixel data and loading initial image data into the input image (lines 11–15). The input image is bound to an Accessor with enabled boundary handling mode *clamping* (lines 18–19). After defining the iteration space, the kernel can be instantiated (line 25) and executed (line 28).

The actual Kernel is implemented by deriving from the framework's provided Kernel base class, inheriting a kernel() method. Within that method the actual kernel code is provided, see Listing 2 (lines 4–7). Because the Gaussian blur filter is a local operator that is performing standard convolution, the kernel can be described using the convolve() method. This method takes three arguments: (a) the mask for defining window size and coefficients; (b) the reduction type; and (c) a C++ lambda function describing the computational steps that should be applied in each iteration. Besides convolve(), HIPA$^{cc}$ offers similar language constructs for local operators to handle reductions (reduce()) and iterations (iterate()) in general.

*2) Generating Code for Vivado HLS:* Considering Vivado HLS as a target for code generation involves numerous challenges to overcome. Convolution masks provided in DSL code must be translated in a more suitable version (integer arithmetic) for FPGAs and hardware accelerators. The same applies to DSL vector types that need to be wrapped into integer streaming buffers for pipelining. In particular, the buffer-wise execution model, where kernels are issued one by one, must be transformed into streaming buffers for pipelining. Hereby, a pipelined structural description is inferred from the linear execution order of kernels. Furthermore, kernel implementations need appropriate placement of Vivado HLS *pragmas* depending on the desired target optimization. This is mostly done by instantiating the right building blocks, encapsulated in a library [14] that is shipped with the generated code.

*B. Optimization Feedback Loop*

In FPGA designs, often more than just a single algorithm has to be placed on one and the same FPGA. Block matching for instance could benefit from a Gaussian blur preprocessing step to increase the likelihood for positive matches, as well as median filtering for postprocessing to eliminate salt and pepper noise. Therefore, often constraints can be defined, such as a resource limitation, in order to ensure that all algorithms fit into the available resources of an FPGA device.

Pragmas set by HIPA$^{cc}$ influence decisively the synthesis results produced by Vivado HLS. Those are mostly affecting the achievable Initiation Interval (II)[3] and resource usage. The II directly impacts the achievable throughput of the algorithm in strong correlation with the clock frequency the synthesis was able to cope with. In fact, the overall latency of an image algorithm can be defined by: #pixels $\times$ $^{II}\!/_{clk.\ freq.}$ plus the initial latency for filling the pipeline, which is negligible for larger image dimensions.

To stay within a given resource budget or to ensure certain timing constraints, in this work, we introduce an optimization feedback loop, which is exploiting the HIPA$^{cc}$ compiler and Vivado HLS. Hereby, synthesis results are analyzed and fed back into the HIPA$^{cc}$ compiler in order to generate a more suitable version. That feedback loop primarily considers three optimization targets: II, clock frequency, which both essentially represent the achievable throughput, and resource usage. For two of those targets, an upper limit can be defined as constraint. The third non-constrained target will serve as a variable parameter, which is iteratively modified by the optimization loop. Early results have shown that exploring different target II's is not a practical approach. For synthesis, always the lowest possible II should be chosen. Otherwise the achievable gain in clock frequency is in most cases not able to keep up with the increased II, which leads to an overall throughput reduction.

The optimization feedback loop attempts to search a suitable version in two phases, as illustrated in Algorithm 1. Initially, the constraints need to be defined, as well as the target type for which a variable parameter is evaluated. In the first phase (line 4–8), that variable parameter is consecutively doubled until all constraints are met. Hereby, the upper bound for the search interval of the second phase is determined. In the second phase (line 9–18), the actual optimization takes place. The search interval is explored by applying the bisection method. Meaning in each iteration, the interval center is chosen as pivot element and represents the upper or lower interval boundary for the next iteration, depending on whether or not the constraints have been met.

---

[3]number of clock cycles a pipelined execution needs to produce an output value, when the pipeline has already been filled

**Algorithm 1** Optimization Feedback Loop

1: **function** OPTIMIZE(target, constraints)
2:   low ← DEFAULTLOW(target)
3:   high ← low
4:   **repeat**                              ▷ Phase 1: Find upper bound
5:     high ← 2×high
6:     GENERATECODE(target, high, constraints)
7:     RUNSYNTHESIS( )
8:   **until** CONSTRAINTSMET(constraints)
9:   **while** low ≠ high **do**            ▷ Phase 2: Search optimum
10:    current ← low+high/2
11:    GENERATECODE(target, current, constraints)
12:    RUNSYNTHESIS( )
13:    **if** CONSTRAINTSMET(constraints) **then**
14:      high ← current
15:    **else**
16:      high ← low
17:    **end if**
18:  **end while**
19: **end function**

### C. The Bit-Count Problem

During the comparison step within Census difference block matching, the Hamming distance needs to be evaluated. Counting bits within various data types can be accomplished fairly efficient in software with the Brian Kernighan Algorithm shown in Listing 3. The number of loop iterations exactly represents the number of set bits to count. As HIPA[cc] supports the use of standard C++, software developers might tend to implement bit counting using this algorithm. Unfortunately, Vivado HLS does not cope with variable loop boundaries and is not able to successfully analyze that the maximum number of iterations solely depends on the bit width of the given data type. As a consequence, unrolling can not be applied, pipelining fails, and no II can be determined.

```
1 int count = 0;
2 while (val) {
3   val &= val - 1;
4   ++count;
5 }
```
Listing 3. Brian Kernighan Algorithm

Whenever falling back to standard C++ code, without enforcing the use of DSL constructs, efficient target-specific code generation might be dramatically limited. This also holds for the above example, which will produce non-pipelined synthesis results. However, implementing the same algorithm with DSL constructs, considering their limitations, forces the developer to introduce a fixed upper bound for the number of iterations. A possible implementation can be seen in Listing 4. Hereby, the early jump (line 3) is maintained for rather fortunate cases and code generation can be applied more tailored to target-specific Vivado HLS characteristics.

The code HIPA[cc] generated specifically for Vivado HLS can be found in Listing 5. The iteration is mapped straight-forward to a loop with static boundaries and additional HLS pragmas have been inserted. With this implementation, pipelining can be enabled and therefore, efficient synthesis results can be achieved.

```
1 int count = 0;
2 iterate(sizeof(val)*8, [&] () {
3   if (!val) break_iterate();
4   val &= val - 1;
5   ++count;
6 });
```
Listing 4. Brian Kernighan Algorithm in DSL Code

```
1 int count = 0;
2 for (int i = 0; i < sizeof(val)*8; ++i) {
3 #pragma HLS unroll
4 #pragma HLS loop_tripcount min=0 max=sizeof(val)*8
5   if (!val) break;
6   val &= val - 1;
7   ++count;
8 }
```
Listing 5. Generated Brian Kernighan Algorithm

## IV. EVALUATION AND RESULTS

Our results for the stereo matching algorithms have been evaluated on the Zynq platform. The algorithms have been implemented in DSL code, which could also be used to target completely different architectures, like GPUs, without any effort. The code used for synthesis by Vivado HLS was generated with HIPA[cc]. To evaluate the quality of the our results, we compare it to handwritten implementations. Furthermore, we will present the results we were able to obtain by applying the optimization feedback loop.

### A. Experimental Environment

**Xilinx Zynq 7100** is a System on Chip (SoC), which tightly integrates an ARM Cortex-A9 dual core CPU and a Kintex FPGA. The included FPGA offers 277,400 Lookup Tables (LUTs), 554,800 flip-flops, 3,020 kB of on-chip memory (BRAM), and 2,020 DSP slices.

**Xilinx Vivado HLS** is a High-Level Synthesis tool specifically targeting Xilinx FPGAs. It allows design entry in C/C++ or SystemC and delivers HDL code (VHDL, Verilog, and SystemC) for synthesizable IP cores. For our experiments we are using the most recent version Vivado HLS 2014.4.

*1) Handwritten Implementation:* In [15] stereo block matching has been realized as a generic VHDL template, which is scalable in several functional and structural parameters like image size, disparity and window block size. By utilizing special buffering techniques it was possible to implement it as streaming architecture, in order to have a direct interface to the image sensor for performing block matching in real time on HD images. For achieving high frame rates, the architecture has been pipelined. Since no specific IP core interfaces have been used, it is easy to port it to a different FPGA vendor or family and may also be base for an ASIC design. The cost functions are calculated by a Processing Element (PE). This common interface allows to switch between different cost functions easily. Depending on the designer constraints (FPGA resources, depth map accuracy) the architecture can be adapted. The minimum detection module MIN has been implemented as a pipelined binary tree. An overall architecture is shown in Figure 4. The images were taken from the Middlebury 2003 stereo datasets [16], which provide several scenes for benchmarking of stereo matching algorithms. The resulting depth maps show different matching qualities depending on the used cost function.

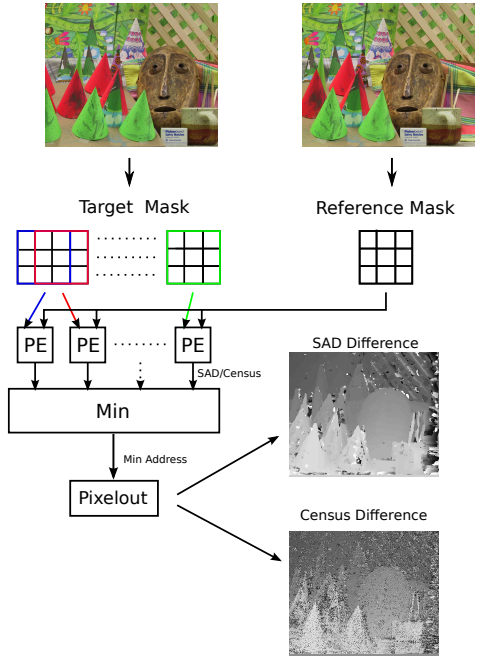

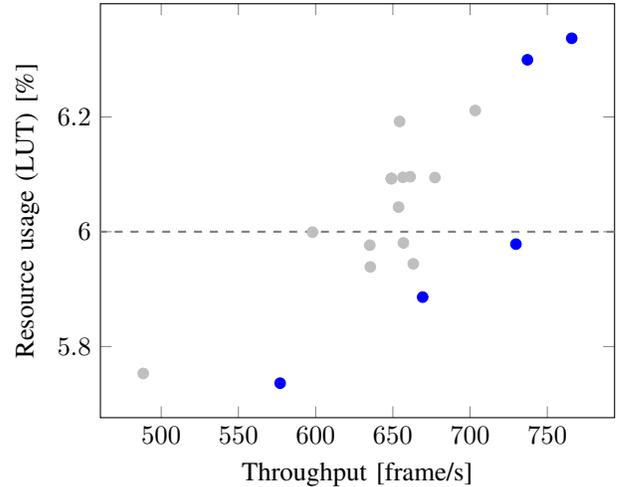

Figure 6. Design points from optimization run with the constraints II = 1 and resource usage < 6 % of the Census block matching algorithm with image size $450 \times 375$ on a Zynq 7100. Blue dots represent Pareto optimal points.

into a streaming pipeline.

*B. Automatic Optimization Results*

Running the evaluation with the optimization feedback loop greatly reduces the number of synthesis runs necessary to converge to predefined constraints. Instead of uniformly investigating the whole search space, the less promising spots are skipped rather early, whereas the most promising spot is very thoroughly explored. Figure 6 shows the results from an optimization run with the constraints II = 1 and resource usage < 6 %. The optimization algorithm is varying the target clock frequency. It can be seen that there are some outliers, which are produced by the synthesis runs at the boundaries of the search space interval. The bisection method enforces synthesis runs close to the constraint (dashed line) rather quickly. Therefore, a cluster forms near that resource constraint line. However, it is not ensured that the last iteration of the run will lead to the best result. As Vivado HLS uses heuristics internally, a slightly lower target clock frequency might lead to a better result than a higher target frequency. For example, in the presented graph it was possible to achieve a clock timing of $8.121$ ns by specifying the desired target clock to $12.735$ ns. On the other hand, the next iteration of the optimization loop resulted in a clock timing of $9.910$ ns when specifying a slightly faster timing of $12.730$ ns. For that reason, a thorough exploration at the constraint boundary is very reasonable. The optimization loop keeps track of all results and reports the one with the highest achieved frequency that is still meeting all defined constraints.

*C. Comparison: HIPA$^{cc}$ vs. Handwritten HDL Code*

A comparison of both algorithm types, generated with HIPA$^{cc}$ and their handwritten equivalents, can be found in Table I. An image size of $450 \times 375$ has been chosen, whereas both implementations are kept generic enough to synthesize accelerators for other image dimensions as well. For the HIPA$^{cc}$ generated implementation, we ran the optimization loop with the constraints II = 1 and resource usage ≤ 100 %. Hereby, we wanted to avoid artifacts introduced by Vivado HLS's internal heuristics, as described above.

Figure 4. Overall structure of the described architecture. Each image of the stereo pair gets streamed through specialized buffers. The PEs have parallel access of all mask values. Each PE gets a sub window from the target mask and all pixel values from the reference mask and perform the defined cost function in parallel. With each clock cycle the minimum module compares the resulting values from the PEs and writes the index of the PE with the lowest value to the output. As a result a depth map from the given surrounding can be calculated.

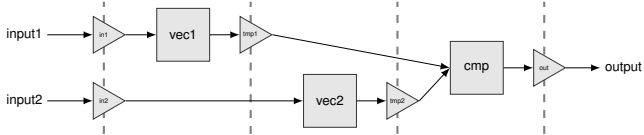

Figure 5. Sequential execution of HIPA$^{cc}$ kernels for computing the Census difference as a pipeline of local operators (squares), Triangles mark buffers and dashed lines represent host barriers between kernel executions.

*2) HIPA$^{cc}$ Implementation:* The DSL implementation for SAD bock matching consists of a single kernel implementing a local operator with two loops. The first for iterating over both windows of the input images. The second for moving the second window along the epipolar line. Code generation is rather straight forward and the quality of the synthesis results almost solely depends on the setting the correct HLS pragmas.

For the Census difference, the process is quite different. Here, the DSL code describes a buffer-wise execution, as shown in Figure 5.Instead of only describing a single kernel, two kernels are necessary. The first one (vec) is a local operator of the window size $5 \times 5$ for computing binary vectors, representing the relation to the surrounding pixels. This kernel is instantiated twice, once for each input image. The second kernel (cmp) is a local operator as well, with a windows size of $60 \times 1$ representing the epipolar line. It compares the binary vectors computed in the previous step within that window and stores the position of the closest match. The intermediate result (the binary vectors) are stored in a temporary buffer. Through code generation, these buffers will be eliminated and replaced by stream objects provided by Vivado, in order to transform the buffer-wise execution model

14

Table I
SYNTHESIS RESULTS OF BLOCK MATCHING ALGORITHMS SAD AND CENSUS DIFFERENCE FOR AN IMAGE OF SIZE $450 \times 375$ ON A ZYNQ 7100.

| | HIPA$^{cc}$ | | | | | | | Handwritten | | | | | | |
|---|---|---|---|---|---|---|---|---|---|---|---|---|---|---|
| | II | LAT | BRAM | DSP | FF | LUT | F[MHz] | II | LAT | BRAM | DSP | FF | LUT | F[MHz] |
| SAD | 1 | 181,797 | 8 | 2 | 140,228 | 66,185 | 182.38 | 1 | 170,565 | 4 | 0 | 29,288 | 37,940 | 271.59 |
| Census | 1 | 180,090 | 8 | 0 | 54,016 | 23,144 | 289.52 | 1 | 170,561 | 4 | 0 | 9,978 | 19,247 | 319.18 |

Vivado HLS was able to achieve an II of 1 for both HIPA$^{cc}$ generated implementations. Therefore, the overall latency of those algorithms is similar compared to the handwritten performance. Regarding resource usage, the number of LUTs is slightly higher (20 %) for the Census difference and up to 74 % higher for SAD. Describing the SAD block matching algorithm in HIPA$^{cc}$ requires language features that are currently not available. This leads to a window size within the local operator that is considerably larger than actually necessary, which can of course be avoided in the handwritten implementation. Due to this deficiency, the achievable clock frequency for SAD is noticeably lower (33 %) compared to the Census difference (9 %). Unfortunately, the number of used flip-flops tremendously exceeds the amount of flip-flops allocated by the respective handwritten equivalent. As the exceedance is similarly large for both, the Census difference and SAD, we attribute this issue to shortcomings within Vivado HLS.

Even though the handwritten implementation is more efficient compared to the version generated by HIPA$^{cc}$, code generation still gives great benefits. First of all, the productiveness is significantly increased, as the necessary lines of DSL code are less than a quarter of the handwritten implementations. Second, the developer does not need to be an FPGA expert. In fact, the DSL code is completely independent of the target architecture. Therefore, the exact same algorithm code can be used to target GPUs or other dedicated accelerators (like the Intel Phi) as well.

## V. CONCLUSION

In this work, we have presented an optimization feedback loop coupled with a DSL compiler. In contrast to handwritten HDL code or even handwritten HLS code, DSLs offer great productivity and deliver fairly good results. Through architecture knowledge provided within the DSL compiler, it is ensured that the generated code variants are efficient target-specific implementations, even though if the developer is not an architecture expert. Despite that, the most important benefit of DSLs is that not only functional portability but also performance portability is provided through those target-specific implementations. However, the most compelling argument for code generation is to easily change large parts of code by just flipping a compiler switch. Therefore, this offers the great possibility to interlock this approach with an automatic optimization loop. This optimization feedback loop can be used for rapid exploration of different code variants given predefined constraints. Therefore, this extension to the existing approach offers further control over code generation and gives developers the possibility to automatically optimize their implementations towards the desired design target without rewriting their code.

## ACKNOWLEDGMENT

This work is supported by the German Research Foundation (DFG), as part of the Research Training Group 1773 "Heterogeneous Image Systems".